\documentclass[aps,prd,twocolumn,showpacs,superscriptaddress,groupedaddress,nofootinbib]{revtex4}
\usepackage{xcolor}
\usepackage{graphicx,color,bm}

\usepackage{amsmath,amssymb}
\usepackage{epstopdf}
\usepackage{ulem}
\usepackage{array}
\usepackage{verbatim}
\usepackage[utf8]{inputenc}
\usepackage{rotating}
\newcolumntype{K}[1]{>{\centering\arraybackslash}m{#1}}

\usepackage[colorlinks,linkcolor=blue,citecolor=blue]{hyperref}

\usepackage{soul}

\newcommand{\orcid}[1]{\href{https://orcid.org/#1}{\,\includegraphics[width=8px]{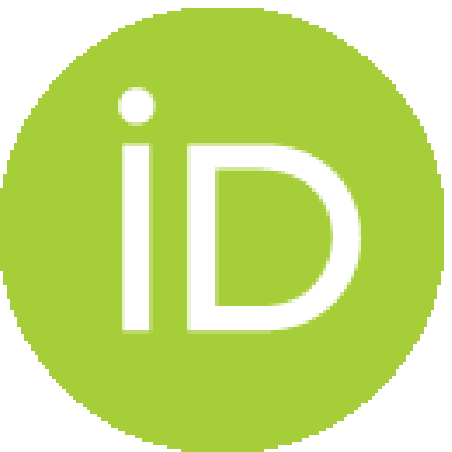}}}

\begin{document}

\title{Constraints on cosmic curvature from cosmic chronometer and quasar observations}

\author{Bikash R. Dinda \orcid{0000-0001-5432-667X}}
\email{bikashdinda.pdf@iiserkol.ac.in}
\affiliation{Department of Physical Sciences, Indian Institute of Science Education and Research Kolkata, Mohanpur, Nadia, West Bengal 741246, India}

\author{Haveesh Singirikonda}
\email{ep17btech11010@iith.ac.in}
\affiliation{Argelander-Institut für Astronomie Auf dem Hügel 71 D-53121 Bonn}

\author{Subhabrata Majumdar}
\email{subha@tifr.res.in}
\affiliation{Department of Theoretical Physics, Tata Institute of Fundamental Research, Dr. Homi Bhabha Road, Navy Nagar, Colaba, Mumbai-400005, India.}

\begin{abstract}
We consider cosmic chronometer (CC) data for the Hubble parameter, quasar (QSO) luminosities data of X-rays and ultraviolet rays emission, and the latest measurements of the present value of the Hubble parameter from 2018 Planck mission (PL18), and SH0ES observations (SHOES) to constrain the present value of cosmic curvature density parameter. We consider three kinds of dark energy models: the $\Lambda$CDM model, the wCDM model, and the CPL parametrization. In all these three models, we find higher values of the matter-energy density parameter, $\Omega_{\rm m0}$ compared to the one obtained from the Planck 2018 mission of CMB observation. Also, we find evidence for a nonflat and closed Universe at 0.5$\sigma$ to 3$\sigma$ confidence levels. The flat Universe is almost 2 to 3$\sigma$, 1 to 1.5$\sigma$, and 0.5 to 1$\sigma$ away from the corresponding mean values, obtained in $\Lambda$CDM model, wCDM model, and CPL parametrization respectively obtained from different combinations of datasets. For example, for the combination of CC and QSO data, we find $\Omega_{\rm k0}=-0.45\pm0.19$, $\Omega_{\rm k0}=-0.36\pm0.24$, and $\Omega_{\rm k0}=-0.145^{+0.215}_{-0.226}$ in the $\Lambda$CDM model, wCDM model, and the CPL parametrization respectively. The evidence for nonzero cosmic curvature is lesser in dynamical dark energy models compared to the $\Lambda$CDM model. That means the evidence of nonzero cosmic curvature depends on the behavior of the equation of state of the dark energy. Since the values of the cosmic curvature are degenerate to the equation of state of the dark energy, we also consider a model independent analysis to constrain the cosmic curvature using the combination of Gaussian process regression analysis and artificial neural networks analysis. In the model independent analysis, we also find evidence for a closed Universe, and the flat Universe is almost 1$\sigma$ away. In the model independent analysis, the constraint on $\Omega_{\rm k0}h^2$ parameter from the combination of CC and QSO data is given as $\Omega_{\rm k0}h^2=-0.30^{+0.34}_{-0.30}$. So, both the model dependent and independent analyses favor a closed Universe from the combinations of CC, QSO, and $H_0$ observations.
\end{abstract}

\keywords{Cosmic curvature, Dark energy, Hubble parameter, quasar, cosmological observations}

\maketitle
\date{\today}

\section{Introduction}
\label{sec-intro}

The possibility of a non-zero spatial curvature, $\Omega_{\rm k0}$ of the Universe is a key question in cosmology. The evidence for it has been tested in many observational tests, through projects like the Planck observation, baryon acoustic oscillation (BAO) measurements, etc \citep{Aghanim:2018eyx,Aiola:2020azj,Beutler:2011hx,Ross:2014qpa,Alam:2016hwk,Scolnic:2017caz,Cooke:2017cwo,Nunes:2020uex,Park:2018tgj,Ryan:2019uor,Collett:2019hrr}. Stronger constraints have been possible in recent years due to advanced technology for telescopes and satellites and improved methods for the analysis of observational data. Theoretically, a spatially flat Universe is predicted by standard inflationary models \citep{1982PhRvL..48.1220A,1987quco.book..149L}. However, it is not trivial to accurately say whether we live in a closed, open, or flat Universe since most measurements leading to estimates of curvature have large error bars and can often support both positive and negative cosmic curvature (inside the 1$\sigma$ region). Also, tensions exist between measurements of curvature by different groups as well; for details see \citep{Handley:2019tkm}. To make it more challenging, a flat Universe exists only for the exact value $\Omega_{\rm k0}=0$, and a slight fluctuation around $\Omega_{\rm k0}=0$ would make our Universe either open or closed. To make it even more complicated, a possible evolving cosmic curvature (called dynamical curvature), i.e, when the Universe can evolve from being open to closed or vice versa, makes it more difficult to get an accurate measurement for the curvature parameter \citep{Desgrange:2019npu}. It has also been shown that local excess mass and energy of the constituents of the Universe can change the value of the curvature in a particular region of space-time. These inhomogeneities are generally averaged over to get a $\Omega_{\rm k0} \approx 0$ \citep{Buchert:1999er,Buchert:2001sa}, but there are concerns that current cosmological analyses ignore the effects of averaging these local inhomogeneities and this could also lead to misinterpretations \citep{Wiltshire:2007fg,Wiltshire:2011vy}. The concept of constant curvature comes directly from the Friedmann Robertson Walker (FRW) metric, where the space-time is also assumed to be homogeneous and isotropic \citep{Coley:2019yov}. Considering all these caveats, it is difficult to accurately estimate our Universe's curvature. The discussion of different issues related to the measurement and interpretation of spatial curvature can be found in reviews like in \citet{Coley:2019yov,DiValentino:2020srs}, and \citet{Coley:2019fpk}.

Currently, a spatially flat Universe is the most widely accepted notion. This has been backed by many observations like the Planck 2018 \citep{Aghanim:2018eyx} and Planck 2015 results \citep{Ade:2015xua} from cosmic microwave background (CMB) observations, baryon acoustic oscillations measurements \citep{Delubac:2014aqe,Ata:2017dya} etc. However several other studies have shown that a non-flat Universe is preferable based on other observations. For example, in \citet{Moresco:2016nqq}, authors put constraints on cosmic curvature using cosmic chronometer data for the Hubble parameter in addition to other data like Planck 2015, where they showed that a slightly negative cosmic curvature is favorable and the flat Universe is 1$\sigma$ away. In \citep{Wei:2016xti,Wang:2017lri,Cheng-Zong:2019iau,Yang:2020bpv}, authors showed that results strongly depend on the different priors on Hubble constant using Hubble data, Supernovae type Ia data, and HII galaxy Hubble diagram. In a different approach, \citet{Li:2019bbg} applied Pade parameterization to the comoving distance and the Hubble parameter and put constraints on the cosmic curvature from standard candles and cosmic chronometer data. They found that the flat Universe may be more than 1$\sigma$ away depending on the Pade parametrization. Recently, \citet{Wang:2019yob} applied the distance sum rule in strong gravitational lensing and supernovae type Ia observations to put model independent constraints on cosmic curvature. They found that a flat or a closed Universe is preferred depending on the different lens models. Also, \citet{Gao:2020irn} examined the connection between cosmic curvature and dark energy and put constraints on these from the latest Pantheon sample. They showed that the open Universe is favorable in wCDM and $w_0 w_a$CDM model at $32\%$ and $78\%$ confidence levels respectively. It is clear that the debate on the value of $\Omega_{\rm k0}$ is far from over.

Most of the studies in the literature put constraints on cosmological parameters (including cosmic curvature) by considering the widely accepted $\Lambda$CDM model. In this benchmark model, the late time acceleration is caused by the presence of the cosmological constant. It is important to consider other dark energy models where the equation of state of the dark energy (eos), $w$ differs from $w=-1$. Due to the presence of strong degeneracy between eos and cosmic curvature density parameter \citep{Clarkson:2007bc,Moresco:2016nqq,Gao:2020irn,Wang:2007mza,Gong:2007wx}, studying evolving dark energy is worth considering when putting constraints on cosmic curvature. In our analysis, we consider the widely used wCDM model and CPL parametrization along with the $\Lambda$CDM model to study the degeneracy between eos and cosmic curvature density parameter.

Degeneracies can lead to possible biases in the estimate of $\Omega_{\rm k0}$, and hence the use of a model independent approach to avoid biases is preferred when putting constraints on cosmic curvature. There have been many efforts to constrain the value of the cosmic curvature density parameter through model independent approaches \citep{Yu:2016gmd,Wei:2019uss,Yang:2020bpv,Liu:2020pfa,Mukherjee:2022ujw}. One of the different approaches is the methodology using the Gaussian process regression (GPR) analysis \citep{NIPS1995_7cce53cf,GpRasWil,Cai:2015pia,Yu:2016gmd,Wang:2017lri,Wang:2019yob,Yang:2020bpv,Liu:2020pfa}. However, most studies of model independent estimation of cosmic curvature tend to assume certain parametrizations, like polynomial expansions, for the luminosity distances and fit the data to it. For example, recently, \citet{Liu:2020pfa} used a third-order logarithmic polynomial to fit the quasar luminosity distance data. It is worth noting at this point that there is no 'a priori' reason to choose a particular parametrization for this fit, and the choice of a parametrization can well impact the conclusions. The path to compute the value of the cosmic curvature starts with the measurements of distances and Hubble parameter at the same redshifts \cite{Clarkson:2007bc}. However, matching redshifts is seldom possible as different surveys focus on different target objects with different redshift distributions. In this regard, GPR has a distinct advantage since it can be used to reconstruct a quantity at a different redshift for which a measurement is not available in the dataset. We also use the GPR in our model independent analysis.

In the present work, we consider three types of observational data to constrain the cosmic curvature density parameter: quasar luminosities (fluxes) data from X-ray and UV ray emissions \citep{Risaliti:2018reu}, cosmic chronometers data for Hubble parameter \citep{Pinho:2018unz,Jimenez_2002}, and the measurement of Hubble constant from Planck 2018 mission \citep{Aghanim:2018eyx} and from SH0ES experiment \citep{Riess:2019cxk}. Further, to study the cosmic curvature density parameter, we consider two approaches: using parametrized dark energy models and considering a model independent analysis. The main aim of this work is to compare the results from the model dependent and independent analyses. In the model independent analysis, we avoid parametrizations like polynomial and Pade (as previously used by others) and we employ a fully non-parametric approach, the Gaussian Process Regression (GPR). However, the GPR uses a mean function which can lead to biases in the results. For this reason, we use a mean function from the artificial neural networks (ANN) which is completely model independent \citep{Wang_2020}. In this way, the combination of GPR and ANN is used in our model independent analysis.

This paper is structured as follows. In Sec.~\ref{sec-basics}, we discuss the basic equations used for calculating cosmological distances. We then describe the Cosmic Chronometers data in Sec.~\ref{sec-chrono}, the luminosity data of quasars in Sec.~\ref{sec-qso}, and the $H_0$ measurements used in Sec.~\ref{sec-H0}. The methodology for the model dependent analysis, with three dark energy models, and the corresponding results are mentioned and summarized in Sec.~\ref{sec-demodels}. In Sec.~\ref{sec:GPR}, we present constraints on the cosmic curvature using model independent analysis. We conclude in Sec.~\ref{sec-conclusion}.

\section{Basics}
\label{sec-basics}

We consider the Friedmann Robertson Walker (FRW) metric including the cosmic curvature term given as

\begin{equation}
dS^{2}= - c^2 a^2 dt^2 + \frac{dr^2}{1-Kr^2} + r^2 d \Omega^2,
\end{equation}

\noindent
where $t$ is the cosmic time, $a$ is the scale factor corresponding to the expansion of the Universe, $r$ is the comoving radial distance, $c$ is the speed of light in the vacuum, and $d \Omega$ is the solid angle element. Here, $K$ is the spatial curvature of the Universe. $K<0$, $K=0$, and $K>0$ correspond to the open, flat, and closed Universe respectively. The line of sight comoving distance, $d_c$, is given as \citep{Hogg:1999ad}

\begin{equation}
    d_c(z) = d_H \int_0^z \frac{dz'}{E(z')} ,
    \label{eq:loscov}
\end{equation}

\noindent
where $z$ and $z'$ correspond to the redshift. $E(z)$ is the normalized Hubble parameter given as $E(z)=H(z)/H_0$, where $H(z)$ is the Hubble parameter and $H_0$ is its present value. Here, $d_H$ is a constant given as $d_{H}=c/H_0$. From the above expression of the line of sight comoving distance, $d_c(z)$, we compute the transverse comoving distance, $D(z)$ given as \citep{Hogg:1999ad}

\begin{equation}
    D(z) = \begin{cases}
    \frac{d_H}{\sqrt{\Omega_{\rm k0}}} \sinh \left( \frac{\sqrt{\Omega_{\rm k0}}}{d_H} d_c(z) \right), & \mbox{if } \Omega_{\rm k0}>0, \\
    d_c(z), & \mbox{if } \Omega_{\rm k0} = 0, \\
    \frac{d_H}{\sqrt{|\Omega_{\rm k0}|}} \sin \left( \frac{\sqrt{|\Omega_{\rm k0}|}}{d_H} d_c(z) \right), & \mbox{if } \Omega_{\rm k0}<0 . \\
    \end{cases}
    \label{eq:trnscov}
\end{equation}

\noindent
$\Omega_{\rm k0}$ is the present value of cosmic curvature density parameter given as $\Omega_{\rm k0}=-K c^{2}/a_0^2H_0^2$, where $a_0$ is the present value of the scale factor.

The luminosity distance, $d_L(z)$ from $D(z)$ is given as \citep{Hogg:1999ad}

\begin{equation}
    d_L(z) = (1+z)D(z).
    \label{eq:lum_dist}
\end{equation}

\section{Observational data and maximum likelihood analysis}
\label{sec-dataset}

We compute bounds on the $\Omega_{k0}$ parameter from the cosmic chronometers data for the Hubble parameter, quasar luminosity data, and the $H_0$ data. Let us briefly discuss these data and the corresponding expressions for the chi-square and loglikelihood.

\subsection{Cosmic chronometers data}
\label{sec-chrono}

In cosmic chronometry, the expansion rate of the Universe i.e. the Hubble parameter is estimated from the measurement of the relative ages of galaxies using spectroscopic dating of the age of the galaxies. This can be seen through the equation, $H(z) = -\frac{1}{1+z}\dfrac{d z}{d t} \approx -\frac{1}{1+z}\frac{\Delta z}{\Delta t}$, where $\Delta t$ is the cosmic time difference between two passively–evolving galaxies that are separated by a small redshift interval $\Delta z$ \citep{Pinho:2018unz,Jimenez_2002}. \footnote{The relative age difference technique is more reliable than the direct determination of absolute ages of galaxies since absolute ages are more vulnerable to systematic uncertainties compared to relative ages.} 

So, in the cosmic chronometers data, we find $z$ vs. $H(z)$ data points (see \citep{Pinho:2018unz,Jimenez_2002} for details of these data points). The corresponding chi-square (denoted by $\chi^2_{\rm CC}$) for these data points is written as

\begin{equation}
    \chi^2_{\rm CC} = \sum_{z} \left[ \frac{H_{\rm obs}(z) - H_{\rm th}(z)}{\Delta H(z)} \right]^2,
    \label{eq:chi_cc}
\end{equation}

\noindent
where $H_{\rm obs}(z)$ is the observed value of the Hubble parameter at a particular redshift and $\Delta H(z)$ is the corresponding $1\sigma$ uncertainty. $H_{\rm th}(z)$ is the theoretical Hubble parameter according to a given model. Throughout this paper, we denote this observation as "CC".

The corresponding loglikelihood, denoted by $\log \mathcal{L}_{\rm CC}$ is given as

\begin{equation}
    \log \mathcal{L}_{\rm CC} = -\frac{\chi^2_{\rm CC}}{2} - \frac{1}{2} \log(|\Sigma_{\rm CC}|) - \frac{n_{\rm CC}}{2} \log(2 \pi),
    \label{eq:lnl_cc}
\end{equation}

\noindent
where $|\Sigma_{\rm CC}|$ is the determinant of the covariance matrix corresponding to the observed uncertainty in the CC data. If there are no covariances between different data points, we can simply have $\log(|\Sigma_{\rm CC}|)=\sum_{z} \log \left( \left[ \Delta H(z) \right]^2 \right)$. $n_{\rm CC}$ is the observed number of data points corresponding to the CC data.

\subsection{The Quasar luminosity data}
\label{sec-qso}

We use emission data of Quasars, obtained in UV and X-rays. These data consist of the luminosities and fluxes of 1598 quasars in the redshift range of $0.036 \leq z \leq 5.1003$. The UV rays are emitted from the gravitationally bound matter in the accretion disk where the matter at high energies emits radiation. In contrast, the X-rays are emitted from hot relativistic electrons and through the Inverse-Compton scattering of UV photons. The quasar luminosity of X-ray emission ($L_X$) and the UV emission ($L_{\rm UV}$) satisfy the scaling relation,

\begin{equation}
    \log_{10} L_X = \gamma \log_{10} L_{\rm UV} + \beta,
    \label{eq:qso_relation}
\end{equation}

\noindent
where $\beta$ and $\gamma$ are two observation related parameters \citep{Avni_1986}. We rewrite the above equation by converting luminosities to the fluxes, $F$ using equation $F=L/4\pi d_{L}^{2}$ given as

\begin{eqnarray}
    \log_{10} F_X &=& \beta + (\gamma-1) \log_{10}(4\pi) + \gamma \log_{10} F_{\rm UV} \nonumber\\
    && + 2 (\gamma -1)  \log_{10} d_L(z),
        \label{eq:qso_relation_2}
\end{eqnarray}

\noindent
where $F_X$ and $F_{\rm UV}$ are the fluxes in X-rays and UV respectively. The parameters $\beta$ and $\gamma$ are treated as independent parameters to be determined from the $\chi^2$ analysis.

For QSO data, the $\chi^2$ (denoted by $\chi^2_{\rm QSO}$) is given as

\begin{equation}
    \chi^2_{\rm QSO} =  \sum_{z} \frac{ \left[ \log_{10} F_X^{\rm obs}(z) - \log_{10} F_{X}^{\rm th}(z) \right]^2}{\sigma^2(z)},
    \label{eq:chi_qso}
\end{equation}

\noindent
where $\sigma^2(z) = \sigma_{X}^2(z) + \delta^2 $. Here, $\log_{10} F_X^{\rm obs}(z)$ is the observed value of the X-ray flux (in $\log_{10}$ scale) and $\sigma_{X}(z)=\Delta \left( \log_{10} F_X^{\rm obs} \right) (z)$ is the corresponding $1\sigma$ error. The theoretical value of the X-ray flux (in $\log_{10}$ scale) for a given UV flux and the luminosity distance calculated using any cosmological model is denoted as $\log_{10} F_X^{\rm th}(z)$. The intrinsic dispersion of the dataset is given by the $\delta$ parameter and it is treated as an independent parameter. Thus, $\beta$, $\gamma$, and $\delta$ are the nuisance parameters for this observation. Throughout this paper, we denote this observation as "QSO".

The corresponding loglikelihood, denoted by $\log \mathcal{L}_{\rm QSO}$ is given as

\begin{equation}
    \log \mathcal{L}_{\rm QSO} = -\frac{\chi^2_{\rm QSO}}{2} - \frac{1}{2} \log(|\Sigma_{\rm QSO}|) - \frac{n_{\rm QSO}}{2} \log(2 \pi),
    \label{eq:lnl_qso}
\end{equation}

\noindent
where $|\Sigma_{\rm QSO}|$ is the determinant of the covariance matrix corresponding to the observed uncertainty in the QSO data. For the diagonal covariance matrix, we get $\log(|\Sigma_{\rm QSO}|)=\sum_{z} \log \left( \sigma^2(z) \right)$. $n_{\rm QSO}$ is the observed number of data points corresponding to the QSO data. Note that, for the QSO data, chi-square minimization is not complete, because the $\delta$ parameter is involved in the second term in the above equation. That is why, for the QSO data, maximum likelihood estimation is more appropriate. For this reason, in this paper, we are doing maximum likelihood analysis instead of chi-square minimization.

\subsection{$H_0$ data}
\label{sec-H0}

For the $H_0$ data, we use the measurements of $H_0$ from the Planck 2018 mission (TT,TE,EE+lowE+lensing) \citep{Aghanim:2018eyx} and SH0ES observation \citep{Riess:2020fzl} alternatively. The $H_0$ values and its $1\sigma$ uncertainties for these two observations are given as

\begin{eqnarray}
H_0 &=& 67.4 \pm 0.5 \hspace{0.05 cm} \left( \text{km} \hspace{0.05 cm} \text{s}^{-1} \hspace{0.05 cm} \text{Mpc}^{-1} \right) \hspace{0.05 cm} \text{(PL18) \citep{Aghanim:2018eyx}},
\label{eq:H0_Planck_18} \\
H_0 &=& 73.2 \pm 1.3 \hspace{0.05 cm} \left( \text{km} \hspace{0.05 cm} \text{s}^{-1} \hspace{0.05 cm} \text{Mpc}^{-1} \right)  \hspace{0.05 cm} \text{(SHOES) \citep{Riess:2020fzl}},
\label{eq:H0_SHOES}
\end{eqnarray}

\noindent
respectively. \footnote{In recent times, the observations of the Hubble constant, $H_{0}$ have received a lot of attention because of the tension between early Universe and late Universe measurements. A good summary of this tension can be found in \citep{Verde}. This is the reason we choose both Planck 2018 (an early time measurement) and SHOES (a late time measurement) results \citep{Dinda:2021ffa}.}

Corresponding to this $H_{0}$ data, the chi-square (denoted by $\chi^2_{\rm H0}$) is written as

\begin{equation}
    \chi^2_{\rm H0} = \left[ \frac{H_0^{\rm obs} - H_0^{\rm th}}{\Delta H_0} \right]^2,
    \label{eq:chi_H0}
\end{equation}

\noindent
where $H_0^{\rm obs}$ is the observed value of $H_0$ and $\Delta H_0$ is the corresponding $1\sigma$ uncertainty. $H_0^{\rm th}$ is the theoretical value of $H_0$ corresponding to a given model. The parameter, $H_0$ is rewritten w.r.t a dimensionless parameter, $h$ given as

\begin{equation}
H_0 = 100 \hspace{0.1 cm} h \hspace{0.1 cm} \text{km s}^{-1}\text{Mpc}^{-1}.
\label{eq:H0th}
\end{equation}

Throughout this paper, we denote Planck 2018 mission (TT,TE,EE+lowE+lensing) and SH0ES observation as "PL18" and "SHOES" respectively.

The corresponding loglikelihood, denoted by $\log \mathcal{L}_{\rm H0}$ is given as

\begin{equation}
    \log \mathcal{L}_{\rm H0} = -\frac{\chi^2_{\rm H0}}{2} - \frac{1}{2} \log \left(2 \pi \left[ \Delta H_0 \right]^2 \right).
    \label{eq:lnl_H0}
\end{equation}

\subsection{Joint analysis}

Given all the above three kinds of data, the total loglikelihood ($\log \mathcal{L}_{\rm tot}$) for a given model is written as

\begin{equation}
\log \mathcal{L}_{\rm tot} = \log \mathcal{L}_{\rm QSO} + \log \mathcal{L}_{\rm CC} + \log \mathcal{L}_{\rm H0}.
\label{eq:lglktot}
\end{equation}

\noindent
Note that, when we need to combine any two of the three datasets, we simply have to add the individual loglikelihoods of the individual dataset, we want to combine.

\begin{figure*}
\includegraphics[width=0.95\textwidth]{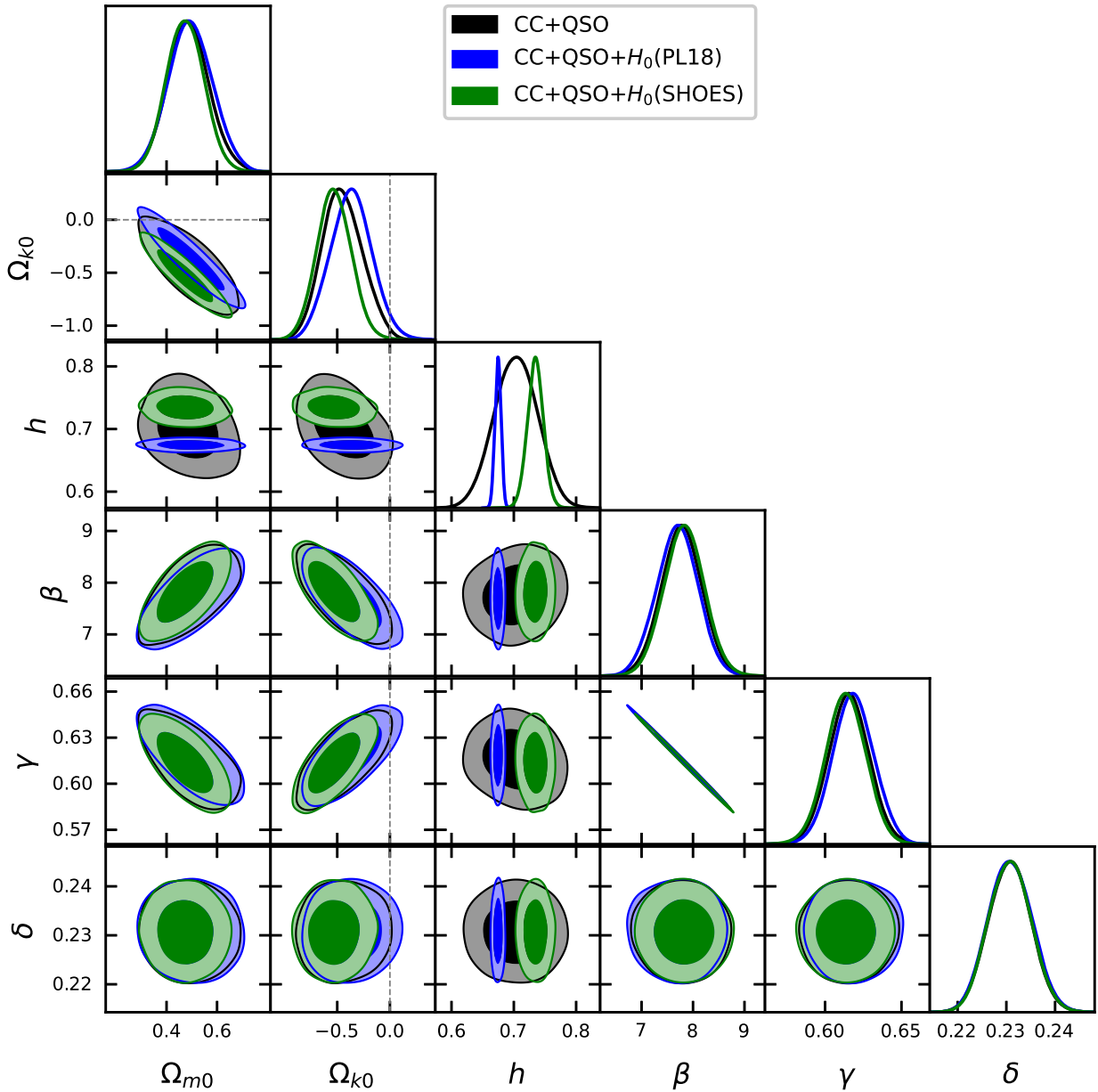}
\caption{Triangle plot (error ellipses between pairs of parameters and 1D marginalized distribution of each parameter) for the parameters of the $\Lambda$CDM model. The black, blue, and green regions are for CC+QSO, CC+QSO+$H_0$(PL18), and CC+QSO+$H_0$(SHOES) combinations of datasets respectively. For a particular color, the darker and the lighter regions correspond to the 1$\sigma$ and 2$\sigma$ regions respectively.}
\label{fig:lcdm_qh}
\end{figure*}

\begin{figure*}
\includegraphics[width=0.95\textwidth]{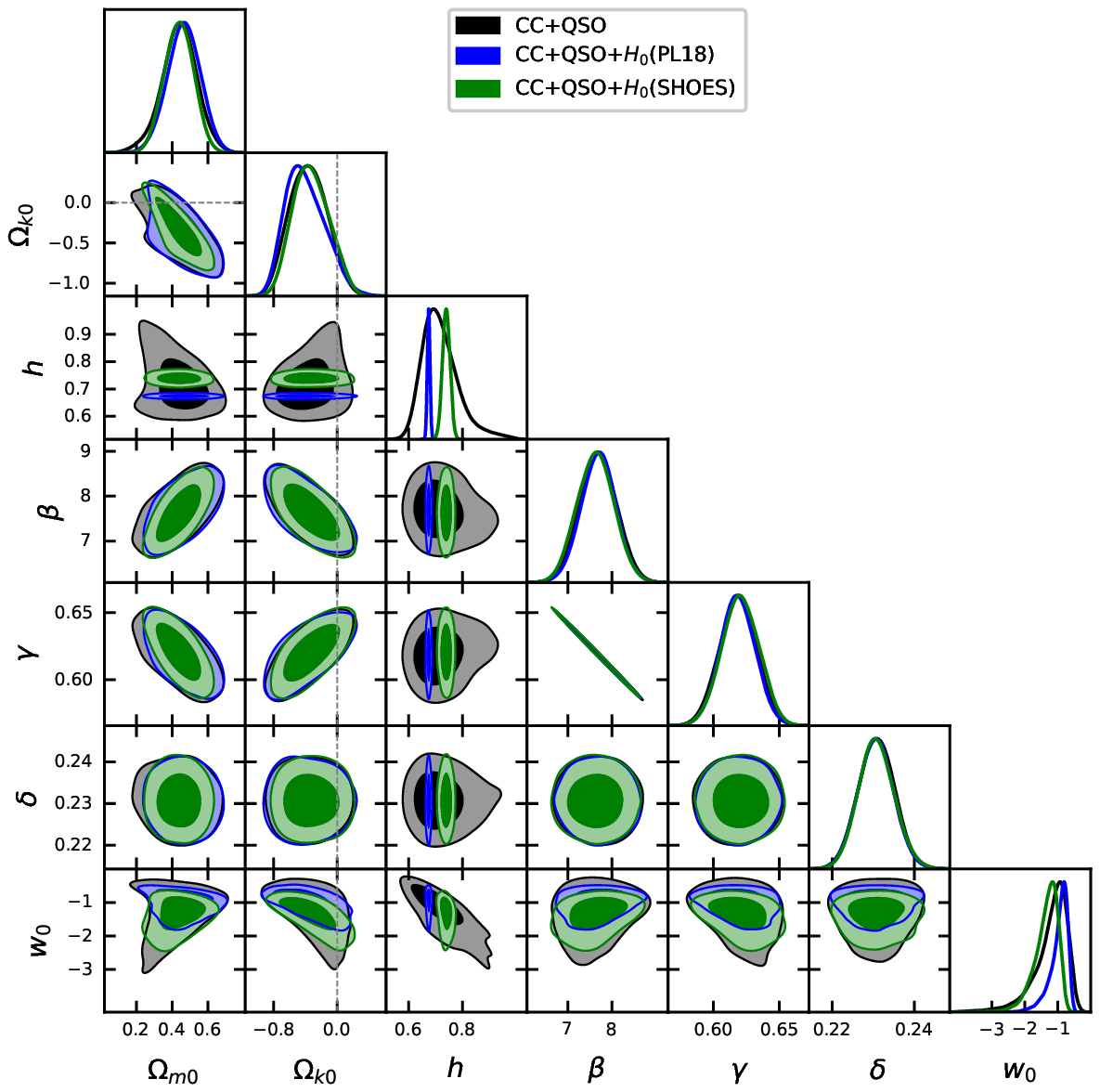}
\caption{Triangle plot for the parameters of the wCDM model. The black, blue, and green regions are for CC+QSO, CC+QSO+$H_0$(PL18), and CC+QSO+$H_0$(SHOES) combinations of datasets respectively. For a particular color, the darker and the lighter regions correspond to the 1$\sigma$ and 2$\sigma$ regions respectively.}
\label{fig:wcdm_qh}
\end{figure*}

\begin{figure*}
\includegraphics[width=0.95\textwidth]{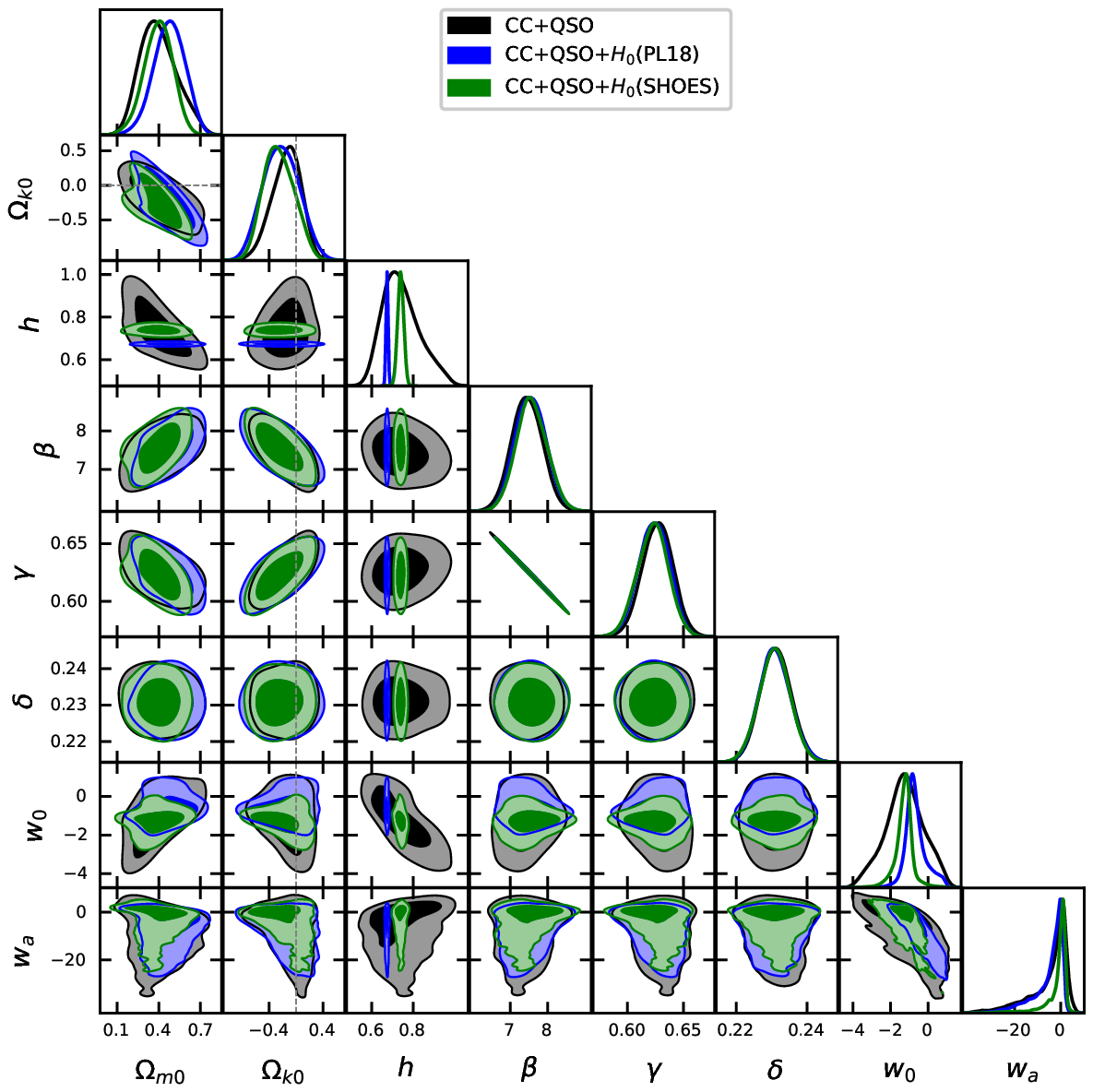}
\caption{Triangle plot for the parameters of the CPL parametrization. The black, blue, and green regions are for CC+QSO, CC+QSO+$H_0$(PL18), and CC+QSO+$H_0$(SHOES) combinations of datasets respectively. For a particular color, the darker and the lighter regions correspond to the 1$\sigma$ and 2$\sigma$ regions respectively.}
\label{fig:cpl_qh}
\end{figure*}

\section{Constraints on $\Omega_{\rm k0}$ in three popular dark energy models}
\label{sec-demodels}

\subsection{Dark energy models}

We consider three widely used dark energy parametrizations where the equation of state of the dark energy (denoted by $w$) is parametrized by some model parameters. These are listed below:

\begin{itemize}
\item
\textbf{$\Lambda$CDM model:} In this model, the equation of state of the dark energy is always $-1$ i.e. $w=-1$ \citep{Aghanim:2018eyx}.
\item
\textbf{wCDM model:} In this model, the equation of state of the dark energy is constant but not fixed to $-1$ \citep{2012ApJ...760...19P,Mortonson:2013zfa,Anselmi:2014nya}. So, we can write the equation of state of the dark energy as $w=w_0$, where $w_0$ is a constant and a parameter of this model. In this way, it is a generalization to the $\Lambda$CDM model.
\item
\textbf{CPL parametrization:} In this model, the equation of state of the dark energy is parameterized as $w = w_0+w_a \frac{z}{1+z}$, where $w_0$ is the equation of state of dark energy at present ($z=0$) and $w_a$ represents the redshift evolution \citep{Chevallier:2000qy,Linder:2002et}. So, it is a generalization to the wCDM model as well as the $\Lambda$CDM model.
\end{itemize}

The wCDM model is the subset of CPL parametrization with one constraint $w_a=0$. $\Lambda$CDM model is the further subset with two constraints $w_0=-1$ and $w_a=0$. In CPL parametrization, $E^2(z)$ is given by

\begin{eqnarray}
&& E^2(z) = \Omega_{\rm m0}(1+z)^3+\Omega_{\rm k0}(1+z)^2 \nonumber\\
&& + \left(1-\Omega_{\rm m0}-\Omega_{\rm k0}\right)(1+z)^{3(1+w_0+w_a)} e^{-\frac{3w_az}{1+z}},
\label{eq:fidE}
\end{eqnarray}

\noindent
where $\Omega_{\rm m0}$ is the present value of the matter-energy density parameter.

Eq.~\eqref{eq:fidE} corresponds to the CPL parametrization. When we put $w_{a}=0$ in this equation, we get the expression corresponding to the wCDM model. Similarly, when we put $w_{a}=0$ and $w_{0}=-1$ in this equation, we get the expression corresponding to the $\Lambda$CDM model. Using the expression of the normalized Hubble parameter from Eq.~\eqref{eq:fidE}, we compute quantities like luminosity distance through Eqs.~\eqref{eq:loscov},~\eqref{eq:trnscov}, and~\eqref{eq:lum_dist}.

\subsection{Maximum likelihood results}

So, now, we have the expression for the Hubble parameter (from Eq.~\eqref{eq:fidE} and using $H(z)=H_0 E(z)$) in a particular model (among the above-mentioned three models). This is the theoretical Hubble parameter, $H_{\rm th}(z)$, which we put in Eq.~\eqref{eq:lnl_cc} to get $\log \mathcal{L}_{\rm CC}$ as a function of model parameters corresponding to the CC data. Note that, in the expression of $H_{\rm th}(z)$, $H_0$ parameter is involved. This parameter, $H_0$ can be seen as $H_0^{\rm th}$ and can be replaced by the dimensionless parameter, $h$ using Eq.~\eqref{eq:H0th}.

Now, knowing the expression of $E(z)$ from Eq.~\eqref{eq:fidE}, we find the luminosity distance in a particular model using Eqs.~\eqref{eq:loscov},~\eqref{eq:trnscov} and~\eqref{eq:lum_dist}. Then, we put this expression of the (theoretical) luminosity distance in Eq.~\eqref{eq:qso_relation_2}, to get $\log_{10} F_{X}^{\rm th}(z)$ and finally, we put this in Eq.~\eqref{eq:lnl_qso} to find $\log \mathcal{L}_{\rm QSO}$ as a function of model parameters. Note that, there are three extra observation-related parameters in $\log \mathcal{L}_{\rm QSO}$. These are $\beta$, $\gamma$ and $\delta$. These can be seen as nuisance parameters.

Now, corresponding to the $H_0$ data, in a particular model, the $\log \mathcal{L}_{\rm H0}$ in Eq.~\eqref{eq:lnl_H0} is trivial to compute since it is simply a function of $h$ by putting Eq.~\eqref{eq:H0th} in Eq.~\eqref{eq:lnl_H0}.

After expressing each loglikelihood, we find loglikelihood corresponding to any particular combination of datasets as functions of the model parameters and the nuisance parameters if QSO data is included. And finally, we minimize the minus of this log-likelihood to get constraints on the model parameters. To do this, we consider the Markov Chain Monte Carlo (MCMC) method. For this, we use the package {\tt emcee} \citep{Foreman_Mackey_2013}. We use the package {\tt GetDist} to draw the corresponding triangle plots for contours \citep{Lewis:2019xzd}.

We present the constraints (from the maximum likelihood analysis) on the model parameters as well as on the nuisance parameters through triangle plots in Figs.~\ref{fig:lcdm_qh},~\ref{fig:wcdm_qh}, and~\ref{fig:cpl_qh} in $\Lambda$CDM model, wCDM model and CPL parametrization respectively. The triangle plots consist of the marginalized 1-dimensional probability distribution of each parameter corresponding to a particular combination of the datasets. These plots also consist of $1\sigma$ and $2\sigma$ contour regions between two parameters for each pair of parameters.

The values of the marginalised $1\sigma$ confidence intervals of each parameter are listed in Tables~\ref{table:ranges_CC_QSO},~\ref{table:ranges_CC_QSO_H0_PL18}, and~\ref{table:ranges_CC_QSO_H0_SHOES} for CC+QSO, CC+QSO+$H_0$(PL18), and CC+QSO+$H_0$(SHOES) combinations of datasets. The important results, we see from these tables (or from the triangle plots), are highlighted below:

\begin{table}[]
    \centering
    \begin{tabular}{|c|c|c|c|}
    \hline
\multicolumn{4}{|c|}{ CC+QSO } \\
    \hline
       Params & $\Lambda$CDM & wCDM & CPL \\ \hline
       $\Omega_{m0}$ & $0.487 \pm 0.082$ & $0.44 \pm 0.10$ & $0.40^{+0.11}_{-0.16}$ \\ \hline
       $\Omega_{k0}$ & $-0.45^{+0.19}_{-0.19}$ & $-0.36^{+0.24}_{-0.24}$ & $-0.145^{+0.215}_{-0.226}$ \\ \hline
       $h$ & $0.703 \pm 0.035$ & $0.718^{+0.046}_{-0.077}$ & $0.741^{+0.067}_{-0.11}$ \\ \hline
       $\beta$ & $7.78 \pm 0.39$ & $7.68 \pm 0.41$ &  $7.45 \pm 0.41$ \\ \hline 
       $\gamma$ & $0.616 \pm 0.013$ & $0.619 \pm 0.014$ & $0.627 \pm 0.014$\\ \hline
       $\delta$ & $0.2307 \pm 0.0043$ & $0.2308 \pm 0.0044$ & $0.2312 \pm 0.0044$ \\ \hline
       $w_{0}$ & - & $-1.20^{+0.59}_{-0.23}$ & $-1.2 \pm 1.0$\\ \hline 
       $w_{a}$ & - & - & $-5.0^{+9.0}_{-2.0}$ \\ \hline
    \end{tabular}
    \caption{Best fit values and 1$\sigma$ ranges of the parameters obtained from the MCMC samples for $\Lambda$CDM, wCDM, and CPL parametrization respectively for CC+QSO combination of datasets.}
    \label{table:ranges_CC_QSO}
\end{table}

\begin{table}[]
    \centering
    \begin{tabular}{|c|c|c|c|}
        \hline
\multicolumn{4}{|c|}{ CC+QSO+$H_{0}$(PL18) } \\
    \hline
       Params & $\Lambda$CDM & wCDM & CPL \\ \hline
       $\Omega_{m0}$ & $0.495\pm 0.087$ & $0.465\pm 0.093$ & $0.47\pm 0.11$ \\ \hline
       $\Omega_{k0}$ & $-0.37^{+0.20}_{-0.20}$ & $-0.40^{+0.26}_{-0.25}$ & $-0.22^{+0.27}_{-0.27}$ \\ \hline
       $h$ & $0.6746\pm 0.0049$ & $0.6740\pm 0.0049$ & $0.6738\pm 0.0051$ \\ \hline
       $\beta$ & $7.69\pm 0.40$ & $7.69\pm 0.39$ &  $7.54\pm 0.41$ \\ \hline 
       $\gamma$ & $0.619\pm 0.013$ & $0.618\pm 0.013$ & $0.624\pm 0.014$\\ \hline
       $\delta$ & $0.2308\pm 0.0044$ & $0.2306\pm 0.0043$ & $0.2309\pm 0.0044$ \\ \hline
       $w_{0}$ & - & $-0.96^{+0.32}_{-0.12}$ & $-0.67^{+0.35}_{-0.58}$\\ \hline 
       $w_{a}$ & - & - & $-5.1^{+6.9}_{-1.3}$ \\ \hline 
    \end{tabular}
    \caption{Best fit values and 1$\sigma$ ranges of the parameters obtained from the MCMC samples for $\Lambda$CDM, wCDM and CPL parametrization respectively for CC+QSO+$H_{0}$(PL18) combination of dataset.}
    \label{table:ranges_CC_QSO_H0_PL18}
\end{table}

\begin{table}[]
    \centering
    \begin{tabular}{|c|c|c|c|}
        \hline
\multicolumn{4}{|c|}{ CC+QSO+$H_{0}$(SHOES) } \\
    \hline
       Params & $\Lambda$CDM & wCDM & CPL \\ \hline
       $\Omega_{m0}$ & $0.474\pm 0.074$ & $0.437\pm 0.081$ & $0.396^{+0.11}_{-0.098}$ \\ \hline
       $\Omega_{k0}$ & $-0.53^{+0.16}_{-0.16}$ & $-0.33^{+0.24}_{-0.23}$ & $-0.25^{+0.24}_{-0.23}$ \\ \hline
       $h$ & $0.735\pm 0.013$ & $0.739\pm 0.014$ & $0.739\pm 0.014$ \\ \hline
       $\beta$ & $7.83\pm 0.39$ & $7.63 \pm 0.42$ &  $7.57\pm 0.41$ \\ \hline 
       $\gamma$ & $0.614\pm 0.013$ & $0.620\pm 0.014$ & $0.623\pm 0.014$\\ \hline
       $\delta$ & $0.2307\pm 0.0043$ & $0.2308\pm 0.0044$ & $0.2308\pm 0.0044$ \\ \hline
       $w_{0}$ & - & $-1.35^{+0.44}_{-0.20}$ & $-1.27^{+0.40}_{-0.32}$\\ \hline 
       $w_{a}$ & - & - & $-1.94^{+4.4}_{-0.042}$ \\ \hline
    \end{tabular}
    \caption{Best fit values and 1$\sigma$ ranges of the parameters obtained from the MCMC samples for $\Lambda$CDM, wCDM and CPL parametrization respectively for CC+QSO+$H_{0}$(SHOES) combination of dataset.}
    \label{table:ranges_CC_QSO_H0_SHOES}
\end{table}

\begin{itemize}
\item
Without an $H_0$ prior, the 1$\sigma$ range of the $H_0$ parameter is broader. When we add the $H_0$ prior (be it from PL18 or SHOES), the 1$\sigma$ range of $H_0$ is narrower according to the prior. One interesting point to notice here is that without $H_0$ priors, i.e. for CC+QSO data, the mean values of $H_0$ are larger compared to the PL18 value, especially, in CPL parametrization. However, the error bars are quite large.
\item
For the $\Omega_{\rm m0}$ parameter, we get interesting results. The mean values of $\Omega_{\rm m0}$ are around 0.4 to 0.5 depending on different models and combinations of datasets. These values are significantly larger than the values obtained from the Planck 2018 mission results \citep{Aghanim:2018eyx}. Further, we see, the mean values of $\Omega_{\rm m0}$ are the largest and lowest for the $\Lambda$CDM model and CPL parametrization respectively for a particular dataset. This fact suggests that by allowing the equation of state of the dark energy to be any constant value (here, corresponding to the wCDM model), we find the mean values of $\Omega_{\rm m0}$ to be smaller compared to the $\Lambda$CDM model. Further, in the case of the evolving equation of state of dark energy (here, corresponding to the CPL parametrization), we find further lower values of $\Omega_{\rm m0}$ compared to the one with a constant equation of state of the dark energy. The 1$\sigma$ ranges of $\Omega_{\rm m0}$ increase from $\Lambda$CDM model to wCDM model and further increase to CPL parametrization. All these facts are true for all three combinations of datasets. The addition of $H_0$ prior does not significantly change the constraints on $\Omega_{\rm m0}$.
\item
We find another interesting result corresponding to the constraints on the $\Omega_{\rm k0}$ parameter. The mean values of $\Omega_{\rm k0}$ are negative for all cases. The flat Universe i.e. $\Omega_{\rm k0}=0$ value is around 2$\sigma$ to 3$\sigma$ away in the $\Lambda$CDM model depending on different combinations of datasets. It is around 1$\sigma$ to 1.5$\sigma$ away in wCDM model. It is around 0.5$\sigma$ to 1$\sigma$ away in CPL parametrization. This means the evidence for a nonflat Universe is strongest in the $\Lambda$CDM model and weakest in the CPL parametrization. This conclusion is not only because of the largest errorbars in $\Omega_{\rm k0}$ constraints in CPL parametrization, it is also because the mean values of $\Omega_{\rm k0}$ are closest to zero for CPL parametrization. This is an interesting result that suggests a strong correlation between cosmic curvature and the behavior of the equation of state of the dark energy.
\item
Both in wCDM model and CPL parametrization, from the mean values and errors of $w_0$ and $w_a$ parameters, we see that the subset $\Lambda$CDM model is inside the 1$\sigma$ regions for all three combinations of datasets. In CPL parametrization, the 1$\sigma$ range in $w_a$ is quite large. This indicates that the datasets, we consider, loosely constrain the evolution of the dark energy equation of state.
\item
The 1$\sigma$ ranges of the nuisance parameter, $\delta$ are almost similar in all three models. That means the constraint on the $\delta$ parameter is insensitive to any model, we consider. Also, it is almost similar in all combinations of datasets. So, the constraint on the delta parameter can be seen as an inherent outcome of the QSO data.
\item
The 1$\sigma$ ranges of $\beta$ and $\gamma$ parameters do not significantly depend on the dark energy models. However, they do depend little on the different combinations of datasets.
\end{itemize}

\section{Model independent analysis using Gaussian process regression and artificial neural networks}
\label{sec:GPR}

We have already seen that the constraints on $\Omega_{\rm k0}$ depend on the behavior of the equation of the state of the dark energy. That means these constraints are model dependent. So, it is necessary to do a model independent analysis to constraint the $\Omega_{\rm k0}$ parameter. To do so, we use Gaussian process regression (GPR) analysis \citep{NIPS1995_7cce53cf,GpRasWil}. The brief details of the GPR are described below:

\subsection{Basics of Gaussian process regression}

The GPR analysis is useful to reconstruct a function, $f$, and the corresponding uncertainty from a dataset. This reconstruction relies on the assumption that each data point satisfies a Gaussian distribution (with mean and standard deviation from the value and error at that data point) and that the whole dataset follows a multivariate Gaussian distribution. The mean and standard deviation of the reconstructed function, $f$ (whose corresponding mean is $\mu (z)$) at a particular point $z$ is determined by a covariance function $cov[f(z),f(\tilde{z})]=k(z,\tilde{z})$ from the data point $\tilde{z}$. This covariance is also called the kernel. Using this kernel, one can generate a vector function $\mathbf{f^*}$ at a set of points (let say $Z^{*}=\{z^{*}_{i}\}$) with $f_{i}^{*}=\mathbf{f^*}(z^{*}_{i})$. So, with the assumption of a Gaussian distribution, this function $\mathbf{f^*}(Z^{*})$ and the corresponding vector function $\mathbf{y}(Z)$ at the observational data points $Z$ can be represented as \citep{NIPS1995_7cce53cf,GpRasWil}

\begin{eqnarray}
\mathbf{f^*}(Z^{*}) &=& \mathcal{N}[\boldsymbol\mu^*,\mathbf{K}(Z^*,Z^*)], \nonumber\\
\mathbf{y}(Z) &=& \mathcal{N}[\boldsymbol\mu,\mathbf{K}(Z,Z)+\mathbf{C}], 
\label{eq:vector_f_Y}
\end{eqnarray}

\noindent
respectively. The symbol $\mathcal{N}$ stands for the Gaussian or normal probability distribution. $\boldsymbol\mu$ and $\mathbf{K}$ are the mean vector and covariance matrix respectively for the corresponding points of $Z$. $\mathbf{C}$ is the covariance matrix of the data. If the data points are uncorrelated (it is the case here), the covariance matrix is simply diagonal as $\mathbf{C} = \{ diag(\sigma_i^2) \}$, where $\sigma_i$ is the standard deviation at each observational data point. Finally, the mean and covariance of the reconstructed function $\mathbf{f^*}$ (at the set of points, $Z^{*}=\{z^{*}_{i}\}$) can be calculated from the data set and this is given as \citep{NIPS1995_7cce53cf,GpRasWil}

\begin{eqnarray}
\left \langle \mathbf{f^*} \right \rangle &=& \boldsymbol\mu^* + \mathbf{K}(Z^*,Z) [\mathbf{K}(Z,Z)+\mathbf{C} ]^{-1}(\mathbf{y-\boldsymbol\mu}), \nonumber\\
cov(\mathbf{f^*},\mathbf{f^*}) &=& \mathbf{K}(Z^*,Z^*) \nonumber\\
&& - \mathbf{K}(Z^*,Z)[\mathbf{K}(Z,Z)+\mathbf{C}]^{-1} \mathbf{K}(Z,Z^*),
\label{eq:mean_cov_f}
\end{eqnarray}

\noindent
respectively. Note that, this approach is sometimes called the {\tt posterior} approach in GPR analysis. In GPR, one important task is to choose the kernel i.e. the covariance function. Here, in our analysis, we chose the squared exponential or Gaussian covariance function given as

\begin{equation}
 k(z,\tilde{z}) = \sigma_f^2 \exp \left[ -\frac{(z-\tilde{z})^2}{2l^2} \right],
 \label{eq:gpr_kernel}
\end{equation}

\noindent
where $\sigma_f$ and $l$ are two hyperparameters that describe the 'bumpiness' of the function. The squared exponential kernel is the popular choice when implementing GPR and is also quite simpler when compared to the other kernels. We also need a mean function, $\mu(z)$ for a GPR analysis. To make the GPR analysis completely cosmological model independent, instead of choosing a mean function from a cosmological model or a zero mean function (which is most popularly used in GPR analysis), we reconstruct a mean function from the data itself using artificial neural networks (ANN) with a popular ANN code {\tt ReFANN} \citep{Wang_2020}.

GPR is also useful to predict the derivatives of the reconstructed functions. For example, the first derivative of the function (denoted by $f'$) and the covariance corresponding to the uncertainty can be predicted at $Z^*$ points in a similar fashion given as \citep{Dinda:2022jih}

\begin{eqnarray}
&& \left \langle \mathbf{f'^*} \right \rangle = \boldsymbol\mu'^* + [\mathbf{K'}(Z,Z^*)]^T [\mathbf{K}(Z,Z)+\mathbf{C} ]^{-1}(\mathbf{y-\boldsymbol\mu}), \nonumber\\
&& cov(\mathbf{f'^*},\mathbf{f'^*}) = \mathbf{K''}(Z^*,Z^*) \nonumber\\
&& - [\mathbf{K'}(Z,Z^*)]^T [\mathbf{K}(Z,Z)+\mathbf{C}]^{-1} \mathbf{K'}(Z,Z^*),
\label{eq:mean_cov_f_derivative}
\end{eqnarray}

\noindent
where we have

\begin{eqnarray}
k'(z,\tilde{z}) &=& \dfrac{\partial k(z,\tilde{z})}{\partial \tilde{z}}, \\
k''(z,\tilde{z}) &=& \dfrac{\partial ^2 k(z,\tilde{z})}{\partial z \partial \tilde{z}}.
\label{eq:kernel_derivatives}
\end{eqnarray}

\noindent
The covariance between the reconstructed function and its first derivative is given as \citep{Dinda:2022jih}

\begin{eqnarray}
&& cov(\mathbf{f^*},\mathbf{f'^*}) = \mathbf{K'}(Z^*,Z^*) \nonumber\\
&& - [\mathbf{K}(Z,Z^*)]^T [\mathbf{K}(Z,Z)+\mathbf{C}]^{-1} \mathbf{K'}(Z,Z^*).
\label{eq:cov_f_fp}
\end{eqnarray}

Finally, to find the kernel hyperparameter values, we minimize the negative of log marginal likelihood given as \citep{NIPS1995_7cce53cf,GpRasWil}

\begin{align}
\label{eq:log_marginal_likelihood}
& \log P(\mathbf{y}|Z) = \nonumber\\
& -\frac{1}{2} (\mathbf{y-\boldsymbol\mu})^T \left[ \mathbf{K}(Z,Z)+\mathbf{C} \right]^{-1} (\mathbf{y-\boldsymbol\mu}) \nonumber\\
& -\frac{1}{2} \log |\mathbf{K}(Z,Z)+\mathbf{C}| -\frac{n}{2} \log{(2 \pi)},
\end{align}

\noindent
where $n$ is the number of observational data points. We obtain the best-fit values of the hyperparameters from the maximum log marginal likelihood analysis using the above equation. We put these best-fit values in Eqs.~\eqref{eq:mean_cov_f},~\eqref{eq:mean_cov_f_derivative}, and~\eqref{eq:cov_f_fp} to get the predictions for the function, its first derivative and the corresponding uncertainties.

\begin{figure*}
\includegraphics[width=0.95\textwidth]{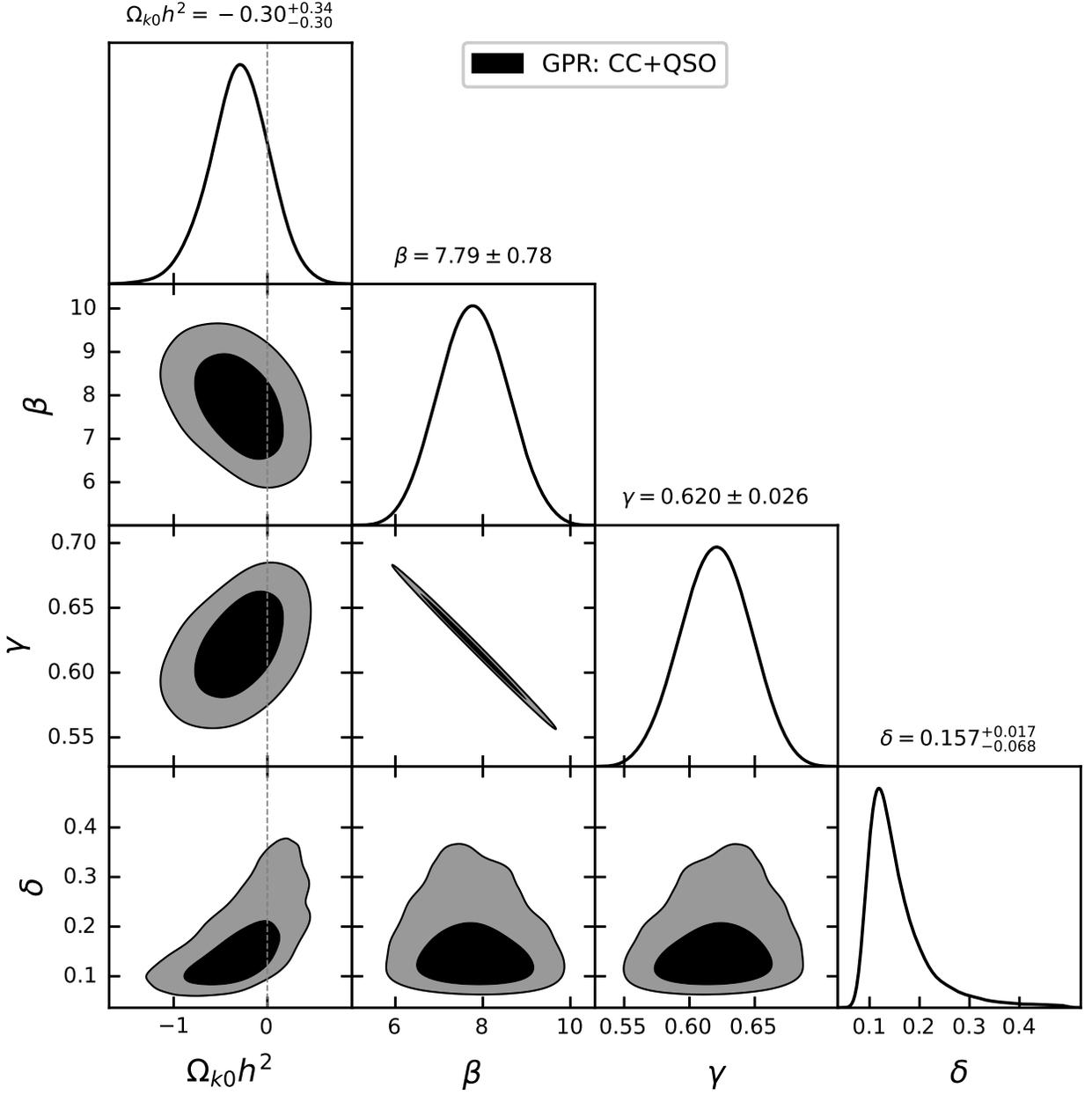}
\caption{Triangle plot for the parameters involved in CC+QSO datasets in model independent analysis. The darker and the lighter regions correspond to the 1$\sigma$ and 2$\sigma$ regions respectively.}
\label{fig:gpr_qh}
\end{figure*}

\begin{figure}
\includegraphics[width=0.45\textwidth]{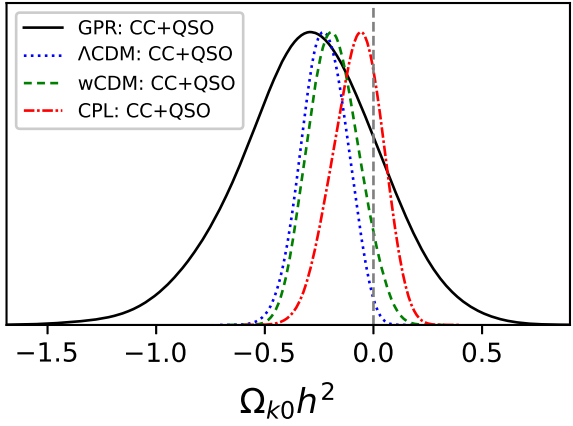}
\caption{Probability distribution of $\Omega_{\rm k0}h^2$ obtained from the maximum likelihood analysis both from model dependent and independent analyses for the CC+QSO combination of datasets. The solid black line corresponds to the model independent analysis. The dotted-blue, dashed-green, and dashed-dotted-red lines correspond to the $\Lambda$CDM model, wCDM model, and CPL parametrization respectively.}
\label{fig:prob_cmp_Ok0h2}
\end{figure}

\begin{figure*}
\includegraphics[width=0.95\textwidth]{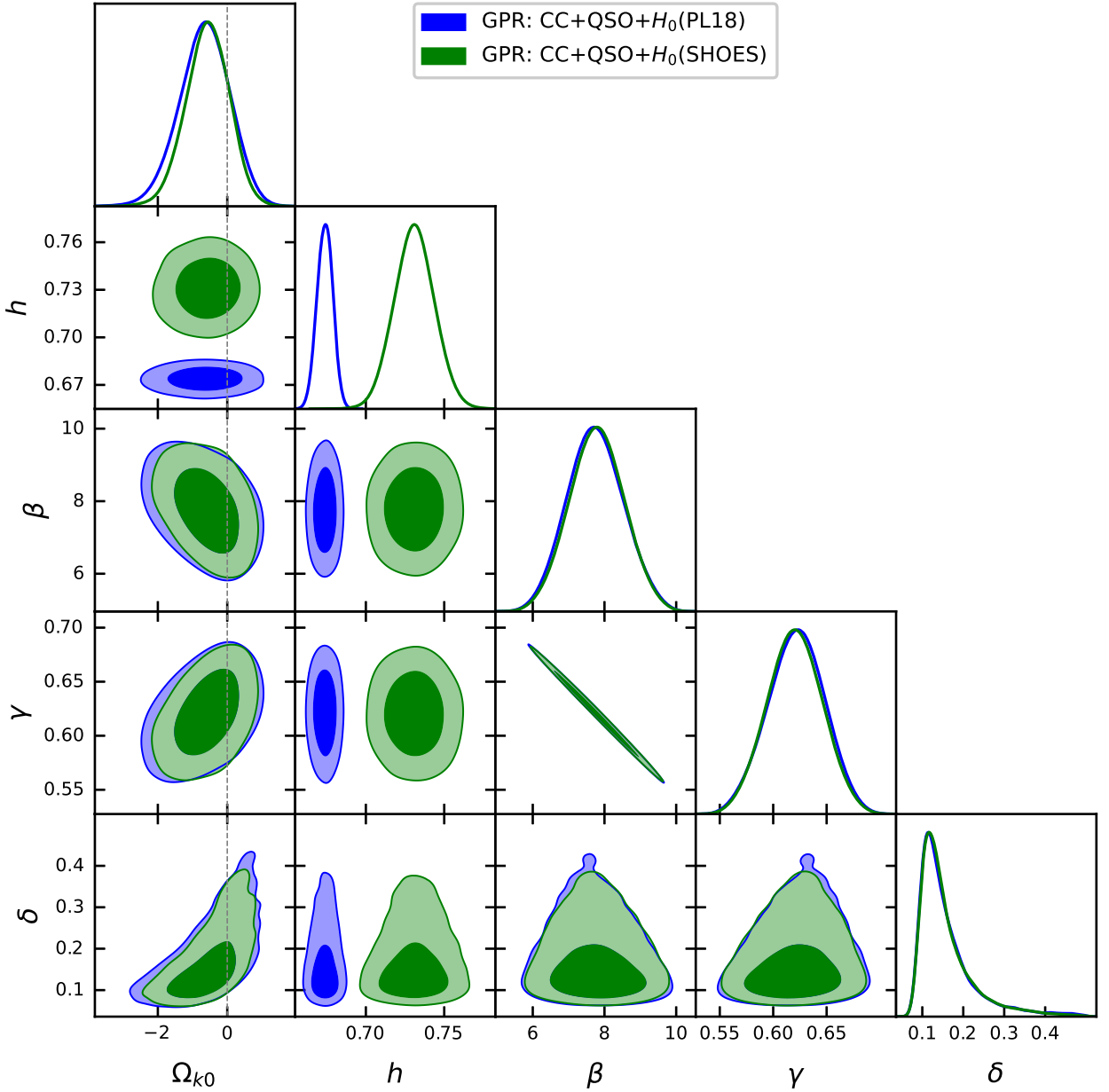}
\caption{Triangle plot for the parameters in model independent analysis. The blue and green regions are for CC+QSO+$H_0$(PL18) and CC+QSO+$H_0$(SHOES) combinations of datasets respectively. For a particular color, the darker and the lighter regions correspond to the 1$\sigma$ and 2$\sigma$ regions respectively.}
\label{fig:gpr_with_h}
\end{figure*}

\begin{figure*}
\includegraphics[width=0.47\textwidth]{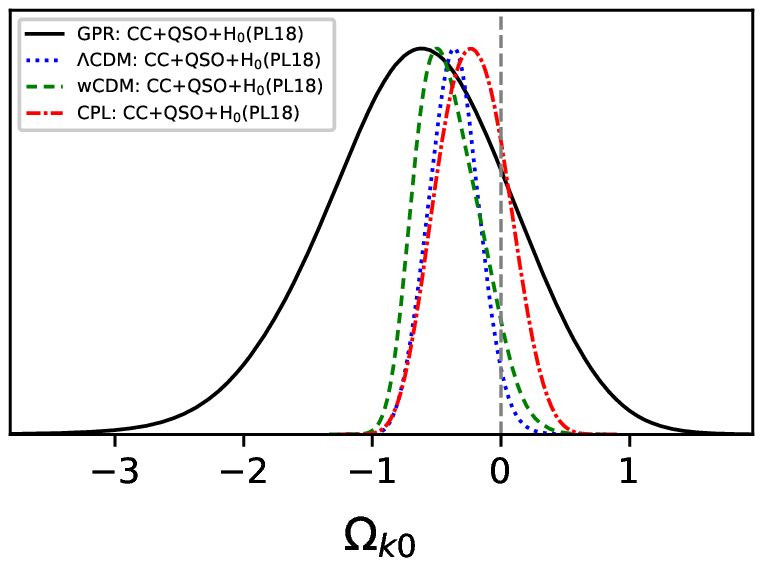}
\includegraphics[width=0.47\textwidth]{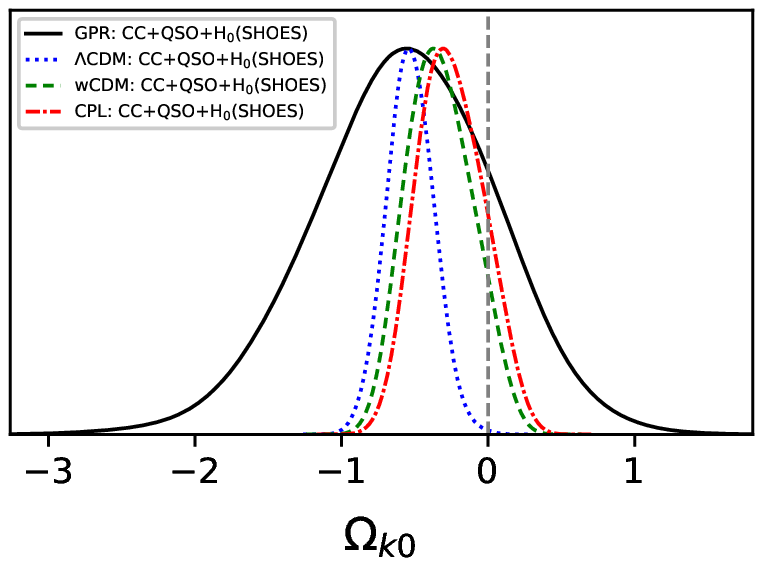}
\caption{Probability distribution of $\Omega_{\rm k0}$ obtained from the maximum likelihood analysis both from model dependent and independent analyses. The left and the right panels correspond to the CC+QSO+$H_0$(PL18) and CC+QSO+$H_0$(SHOES) combinations of datasets respectively. The solid black lines correspond to the model independent analysis. The dotted-blue, dashed-green, and dashed-dotted-red lines correspond to the $\Lambda$CDM model, wCDM model, and CPL parametrization respectively.}
\label{fig:prob_Ok0_cmp}
\end{figure*}

\subsection{Constraints on $(\Omega_{k0}h^2)$ and $\Omega_{k0}$}

To get constraints on the $\Omega_{k0}$ parameter, we compare CC data and QSO data. We can either reconstruct a function for the luminosity distance from the CC data and compare these to the QSO data or we can reconstruct a function for the Hubble parameter from the QSO data and compare these to the CC data. The first one is difficult because of two main reasons. One is as follows: the two datasets are at different redshifts. So, we have to fix one set of redshift points for the entire analysis. Since the QSO data has a larger range of redshift points ($0.036 \leq z \leq 5.1003$) compared to the one for CC data ($0.07 \leq z \leq 1.965$), we fix the redshift points of our analysis to redshifts of the CC data. Either case is not useful because in that case, one has to consider extrapolation of the data set and this would lead to huge error bars in the analysis.

The second reason is even more important as follows: the reconstruction of luminosity distance from the CC data requires an integration through Eq.~\eqref{eq:loscov} and the propagation of uncertainty through the integration equation is difficult to compute. Also, in this case, we can not use GPR analysis because GPR does not predict the integration of a function. On the other hand, GPR can predict the derivatives of a function and the corresponding uncertainties which we have seen in the previous subsection. So, we reconstruct $H(z)$ from the QSO data and compare these with CC data to get constraints on the cosmic curvature. To do this, we have to write the Hubble parameter w.r.t the luminosity distance given as \citep{Dinda:2022vmb}

\begin{equation}
H^2 = \frac{(1+z)^2\left[c^2(1+z)^2+W_{\rm k0}d_L^2\right]}{\left[(1+z)d'_L-d_L\right]^2} ,
\label{eq:Hsqr_wrt_dL_dLp}
\end{equation}

\noindent
where $W_{\rm k0}$ is given as

\begin{equation}
W_{\rm k0} = \Omega_{\rm k0}H_0^2 = 10^4 \left(\Omega_{\rm k0}h^2\right) \hspace{0.05 cm} \left( \text{km} \hspace{0.05 cm} \text{s}^{-1} \hspace{0.05 cm} \text{Mpc}^{-1} \right) .
\end{equation}

\noindent
Everywhere, prime denotes the first derivative w.r.t redshift.

We first use the ANN analysis with the help of {\tt ReFANN} code to reconstruct a mean function for $\log_{10}{F_X}$ from the QSO data. We use this reconstructed mean function in the GPR analysis and we find mean values of $\log_{10}{F_X}$, its first derivative, and the corresponding uncertainties at CC redshift points. For $\log_{10}{F_{\rm UV}}$, we do not need GPR analysis, because these data have no observational error bars. So, we simply use ANN analysis, to reconstruct mean values of $\log_{10}{F_{\rm UV}}$ and its first derivative at the CC redshift points. Using these values, we compute $d_L$, $d'_L$, and the corresponding uncertainties at CC redshift points. This step is straightforward using equations

\begin{eqnarray}
d_L &=& 10^{ \frac{P-\beta-(\gamma-1)\log_{10}{(4\pi)}-\gamma Q}{2(\gamma-1)} }, \\
d'_L &=& \frac{(P'-\gamma Q')d_L \ln{10}}{2(\gamma-1)},
\label{eq:dL_dLp_QSO}
\end{eqnarray}

\noindent
where $P=\log_{10}{F_X}$ and $Q=\log_{10}{F_{\rm UV}}$. We find the corresponding uncertainties using simple propagation of uncertainties through the above equations. Note that $d_L$ and $d'_L$ are functions of $\beta$ and $\gamma$ parameters. The uncertainties in these are also dependent on $\beta$ and $\gamma$ parameters. Also, note that the $\delta$ parameter is involved in the uncertainties in a similar way as in the model dependent case. Once we have $d_L$, $d'_L$, and the corresponding uncertainties, we use these to reconstruct $H(z)$ and the corresponding uncertainties at CC redshift points using Eq.~\eqref{eq:Hsqr_wrt_dL_dLp} and using the corresponding propagation of uncertainties. Note that, the reconstructed $H(z)$ is function of $\beta$, $\gamma$ and $\Omega_{\rm k0}h^2$ parameters. The corresponding uncertainty in the $H(z)$ is function of $\beta$, $\gamma$, $\Omega_{\rm k0}h^2$ and $\delta$ parameters.

We compare the reconstructed $H(z)$ from the QSO data with the observed $H(z)$ from the CC data and denote a corresponding loglikelihood given as

\begin{eqnarray}
&& \log{\mathcal{L}}_{\rm CC+QSO} (\Omega_{\rm k0}h^2,\beta,\gamma,\delta) \nonumber\\
&& = - \frac{1}{2} \sum_{z_{\rm CC}} \frac{ \left[ H_{\rm R}(z_{\rm CC},\Omega_{\rm k0}h^2,\beta,\gamma)-H_{\rm CC} (z_{\rm CC}) \right]^2}{ \Delta H_{\rm tot}^2(z_{\rm CC},\Omega_{\rm k0}h^2,\beta,\gamma,\delta) } \nonumber\\
&& - \frac{1}{2} \sum_{z_{\rm CC}} \log{ \left[ 2 \pi \Delta H_{\rm tot}^2(z_{\rm CC},\Omega_{\rm k0}h^2,\beta,\gamma,\delta) \right] },
\label{eq:lnlk_CC_QSO_GPR}
\end{eqnarray}

\noindent
where we have denoted the reconstructed Hubble parameter as $H_R$ and the observed Hubble parameter from CC data as $H_{\rm CC}$. The $z_{\rm CC}$ corresponds to each redshift point of the CC data. $\Delta H_{\rm tot}^2$ is the total variance in $H$ that arises from both the reconstruction and the CC data given as

\begin{eqnarray}
\Delta H_{\rm tot}^2(z_{\rm CC},\Omega_{\rm k0}h^2,\beta,\gamma,\delta) &=& \Delta H_R^2(z_{\rm CC},\Omega_{\rm k0}h^2,\beta,\gamma,\delta) \nonumber\\
&& +\Delta H_{\rm CC}^2(z_{\rm CC}) ,
\end{eqnarray}

\noindent
where subscripts "R" and "CC" correspond to the reconstruction and the CC data respectively.

We do the maximum loglikelihood analysis using Eq.~\eqref{eq:lnlk_CC_QSO_GPR} to get simultaneous constraints on $\Omega_{\rm k0}h^2,\beta,\gamma$, and $\delta$ parameters. Note that, we can not get constraints on the $\Omega_{k0}$ or $h$ individually in this analysis from only CC and QSO data. The constraints are shown in Fig.~\ref{fig:gpr_qh}. The mean values and the 1$\sigma$ ranges are also mentioned in this figure.

\begin{itemize}
\item
We find the mean value of $\Omega_{\rm k0}h^2$ is negative and the flat Universe is almost 1$\sigma$ away.
\item
The mean values of the $\beta$ and $\gamma$ parameters are similar as in the case for the three dark energy models, but the 1$\sigma$ ranges of these parameters are comparatively large.
\item
The mean value of the $\delta$ parameter is smaller compared to the three dark energy models. The reason is as follows: in the model dependent analysis, directly we have used the QSO data which are intrinsically scattered giving rise to a particular positive value of $\delta$. On the other hand, in the model independent analysis, we reconstruct smooth functions using GPR analysis. This reduces the scatteredness in the reconstructed functional values compared to the original QSO data. Thus we find a lower value of delta in the model independent analysis. However, the error bar in it is comparatively very large.
\end{itemize}

Since we have no constraints on $\Omega_{k0}$ parameter from only CC and QSO data from model independent analysis, we compute constraints on $\Omega_{\rm k0}h^2$ in three dark energy models from constraints on $\Omega_{\rm k0}$ and $h$ to compare the model dependent and independent analysis for the cosmic curvature. In Fig.~\ref{fig:prob_cmp_Ok0h2}, we compare the probability distribution functions of $\Omega_{\rm k0}h^2$ obtained both from model dependent and independent analysis for CC+QSO data. We also mention the mean values and the 1$\sigma$ ranges in Table~\ref{table:ranges_CC_QSO_GPR}.

\begin{itemize}
\item
We find the mean value of $\Omega_{\rm k0}h^2$ is more negative in model independent analysis compared to the three dark energy models. However, the 1$\sigma$ range is comparatively very large in model independent analysis.
\end{itemize}

\begin{table}[]
    \centering
    \begin{tabular}{|c|c|}
        \hline
\multicolumn{2}{|c|}{ CC+QSO } \\
    \hline
       & $\Omega_{\rm k0}h^2$ \\ \hline
       GPR & $-0.30^{+0.34}_{-0.30}$ \\ \hline
       $\Lambda$CDM & $-0.23^{+0.10}_{-0.10}$ \\ \hline
       wCDM & $-0.18^{+0.11}_{-0.12}$ \\ \hline
       CPL & $-0.07^{+0.12}_{-0.11}$ \\ \hline
    \end{tabular}
    \caption{Best fit values and 1$\sigma$ ranges of the $\Omega_{\rm k0}h^2$ parameter for CC+QSO combination of datasets.}
    \label{table:ranges_CC_QSO_GPR}
\end{table}

\noindent
With the addition of $H_0$ prior, we get constraints on the $\Omega_{\rm k0}$ parameter in the model independent analysis. In Fig.~\ref{fig:gpr_with_h}, we show the constraints on the parameters for for CC+QSO+$H_{0}$(PL18) and CC+QSO+$H_{0}$(SHOES) combinations of datasets. The addition of $H_0$ prior in model independent analysis does not change the constraints on $\beta$, $\gamma$, and $\delta$ parameters. It only gives constraints on $\Omega_{\rm k0}$ from constraints on $\Omega_{\rm k0}h^2$.

We compare the probability distribution of $\Omega_{\rm k0}$ obtained both from model dependent and independent analyses in Fig.~\ref{fig:prob_Ok0_cmp}. The left and right panels are for CC+QSO+$H_{0}$(PL18) and CC+QSO+$H_{0}$(SHOES) combinations of datasets respectively. We also mention the mean values and the corresponding 1$\sigma$ ranges in Tables~\ref{table:ranges_CC_QSO_PL18_GPR} and~\ref{table:ranges_CC_QSO_SHOES_GPR} for CC+QSO+$H_{0}$(PL18) and CC+QSO+$H_{0}$(SHOES) respectively.

\begin{table}[]
    \centering
    \begin{tabular}{|c|c|}
        \hline
\multicolumn{2}{|c|}{ CC+QSO+$H_{0}$(PL18) } \\
    \hline
       & $\Omega_{\rm k0}$ \\ \hline
       GPR & $-0.66^{+0.76}_{-0.68}$ \\ \hline
       $\Lambda$CDM & $-0.37^{+0.20}_{-0.20}$ \\ \hline
       wCDM & $-0.40^{+0.26}_{-0.25}$ \\ \hline
       CPL & $-0.22^{+0.27}_{-0.27}$ \\ \hline
    \end{tabular}
    \caption{Best fit values and 1$\sigma$ ranges of the $\Omega_{\rm k0}$ parameter for CC+QSO+$H_{0}$(PL18) combination of datasets.}
    \label{table:ranges_CC_QSO_PL18_GPR}
\end{table}

\begin{table}[]
    \centering
    \begin{tabular}{|c|c|}
        \hline
\multicolumn{2}{|c|}{ CC+QSO+$H_{0}$(SHOES) } \\
    \hline
       & $\Omega_{\rm k0}$ \\ \hline
       GPR & $-0.57^{+0.65}_{-0.58}$ \\ \hline
       $\Lambda$CDM & $-0.53^{+0.16}_{-0.16}$ \\ \hline
       wCDM & $-0.33^{+0.24}_{-0.23}$ \\ \hline
       CPL & $-0.25^{+0.24}_{-0.23}$ \\ \hline
    \end{tabular}
    \caption{Best fit values and 1$\sigma$ ranges of the $\Omega_{\rm k0}$ parameter for CC+QSO+$H_{0}$(SHOES) combination of datasets.}
    \label{table:ranges_CC_QSO_SHOES_GPR}
\end{table}

\begin{itemize}
\item
Similar to the case of $\Omega_{\rm k0}h^2$, we find the mean values of $\Omega_{\rm k0}$ are more negative in model independent analysis compared to the analysis in three dark energy models. The 1$\sigma$ ranges of $\Omega_{\rm k0}$ are very large in model independent analysis compared to the model dependent analysis. Interestingly, we find the nonflat Universe is almost 1$\sigma$ away in the model independent analysis. The dependence of this result on the prior of $H_0$ is not very significant.
\end{itemize}

\section{Conclusions}
\label{sec-conclusion}

We consider three kinds of data to constrain the cosmic curvature density parameter, $\Omega_{\rm k0}$ and these are quasar luminosities data of X-rays and UV rays emission, cosmic chronometers data for the Hubble parameter and the measurement of $H_0$ from Planck 2018 mission and SH0ES experiment.

To compute $\Omega_{\rm k0}$, we first use a model dependent analysis by considering the three popular classes of dark energy models, $\Lambda$CDM, wCDM, and the CPL parametrization.

In all these three dark energy models, we find mean values of $H_0$ to be larger in CC+QSO data compared to the value corresponding to the Planck 2018 mission (PL18).

Interestingly, we find $\Omega_{\rm m0}$ values are around 0.4 to 0.5 depending on the different combinations of datasets. These values are significantly larger compared to the ones in Planck 2018 results.

All these three models suggest a non-flat Universe at different confidence levels. The mean values of the cosmic curvature density parameter are negative. The flat Universe, $\Omega_{\rm k0}=0$ is almost 2 to 3$\sigma$, 1 to 1.5$\sigma$, and 0.5 to 1$\sigma$ away from the corresponding mean values in $\Lambda$CDM, wCDM, and CPL parametrization respectively.

This means the evidence for non-zero $\Omega_{\rm k0}$ strongly depends on the behavior of the dark energy equation of state (eos). This evidence decreases from a fixed eos ($-1$ in $\Lambda$CDM) to a varying but constant eos ($w$=constant including $-1$ in wCDM). It further decreases for evolving eos ($w$ varies with redshift) as in CPL parametrization.

Since the cosmic curvature density parameter is degenerate to the behavior of the equation of state of the dark energy, we also consider a model independent analysis to compute the cosmic curvature density parameter using the combination of Gaussian process regression and artificial neural networks.

Interestingly, in the model independent analysis, we also find that the flat Universe is almost 1$\sigma$ away from the corresponding mean values which are negative.

In summary, we find that the closed nonflat Universe is favorable from quasar luminosities data of X-rays and UV rays, cosmic chronometers data for $H(z)$, and the measurement of $H_{0}$ from the 2018 Planck mission or SH0ES experiment. The flat Universe is 0.5$\sigma$ to 3$\sigma$ away in model dependent analysis and almost 1$\sigma$ away in model independent analysis from the corresponding mean values.


\bibliographystyle{apsrev4-1}
\bibliography{refDEquasar}

\end{document}